\documentclass[fleqn,twoside,twocolumn,nofootinbib,showkeys]{revtex4} % Specifies the document class %,unsortedaddress
\usepackage[sec,nocpr]{ujp} % \usepackage[cyr]{ujp} for cyrillic
\begin{document}
\title[Searching for the QCD critical point with net-proton number fluctuations]%колонтитул
{SEARCHING FOR THE QCD CRITICAL POINT WITH NET-PROTON NUMBER FLUCTUATIONS}%
\author{Micha\l\  Szyma\'{n}ski}%1
\affiliation{Institute of Theoretical Physics, University of Wroc\l aw}%институт 1
\address{PL-50204 Wroc\l aw, Poland}%адрес 1
\email{michal.szymanski@ift.uni.wroc.pl}%e-mail 1

\author{Marcus Bluhm}%2
\affiliation{Institute of Theoretical Physics, University of Wroc\l aw}%институт 1
\address{PL-50204 Wroc\l aw, Poland}%адрес 1
\affiliation{~SUBATECH UMR 6457 (IMT Atlantique, Universit\'e de Nantes, IN2P3/CNRS)}%институт 2
\address{4 rue Alfred Kastler, 44307 Nantes, France}%адрес 2
\affiliation{Extreme Matter Institute EMMI, GSI}
\address{Planckstr. 1, D-64291 Darmstadt, Germany}

\author{Krzysztof Redlich}%2
\affiliation{Institute of Theoretical Physics, University of Wroc\l aw}%институт 1
\address{PL-50204 Wroc\l aw, Poland}%адрес 1
\affiliation{Extreme Matter Institute EMMI, GSI}
\address{Planckstr. 1, D-64291 Darmstadt, Germany}

\author{Chihiro Sasaki}%2
\affiliation{Institute of Theoretical Physics, University of Wroc\l aw}%институт 1
\address{PL-50204 Wroc\l aw, Poland}%адрес 1

\udk{539}
\razd{\seci}

\autorcol{M.\hspace*{0.7mm}Szyma\'{n}ski, M.\hspace*{0.7mm}Bluhm, K.\hspace*{0.7mm}Redlich, C.\hspace*{0.7mm}Sasaki}%

\setcounter{page}{1}%

\begin{abstract}
Net-proton number fluctuations can be measured experimentally and hence provide a source of important information about the matter created during relativistic heavy ion collisions. Particularly, they may give us clues about the conjectured QCD critical point. In this work the beam-energy dependence of
ratios of the first four cumulants of the net-proton number is discussed. These quantities are calculated using
a phenomenologically motivated model in which critical mode
fluctuations couple to protons and anti-protons. Our model
qualitatively captures both the monotonic behavior of the
lowest-order ratio as well as the non-monotonic behavior of higher-order ratios, as seen in the experimental data from the STAR Collaboration. We also discuss the dependence of our results on the coupling strength and the location of the critical point.
\end{abstract}


\maketitle

\section{Introduction}

The theoretical and experimental investigation of the phase diagram of strongly interacting matter is an  important subject of modern high energy physics. One of the unresolved questions concerns the existence and location of the QCD critical point (CP) in the $T$ and $\mu$ plane. Strong fluctuations of the critical mode, $\sigma$, in the vicinity of CP, although not directly observable, are expected to couple to physically measurable quantities, such as fluctuations of conserved charges~\cite{Stephanov:1998dy,Stephanov:1999zu}.

Fluctuations of the net-proton number serve as an experimental probe of baryon number fluctuations. Recent, but still preliminary results of the STAR Collaboration~\cite{Luo:2015ewa,Luo:2015doi,Thader:2016gpa} show a non-monotonic beam energy dependence of the ratios of higher order net-proton number cumulants. However, the interpretation of the data is still unclear~\cite{T4,T5,V1,V2} and therefore effective models are needed to improve our understanding of these quantities.\\
\indent One of such models was developed in~\cite{Bluhm:2016byc}, where the impact of resonance decays on net-proton number cumulant ratios was studied. This model could qualitatively describe the non-monotonic behavior of the $C_3/C_2$ and $C_4/C_2$ ratios. However, it also showed a strong non-monotonic behavior of the $C_2/C_1$ ratio which is not observed experimentally. Recently, this model was re-examined~\cite{Szymanski:2019yho} to take into account the scaling properties of the baryon number and chiral susceptibilities obtained within effective models~\cite{Sasaki:2006ws,Sasaki:2007qh}. This reduces the effect of critical fluctuations in the net-proton number variance and thus allows for a better description of the STAR data.

Here we discuss the beam energy dependence of the ratios of net-proton number cumulants obtained using the refined model from Ref.~\cite{Szymanski:2019yho} and study their dependence on the coupling strength between critical mode and (anti)protons as well as their dependence on the location of the critical point.

\section{Model setup}

\begin{figure}
\vskip1mm
  \includegraphics[width=0.85\linewidth]{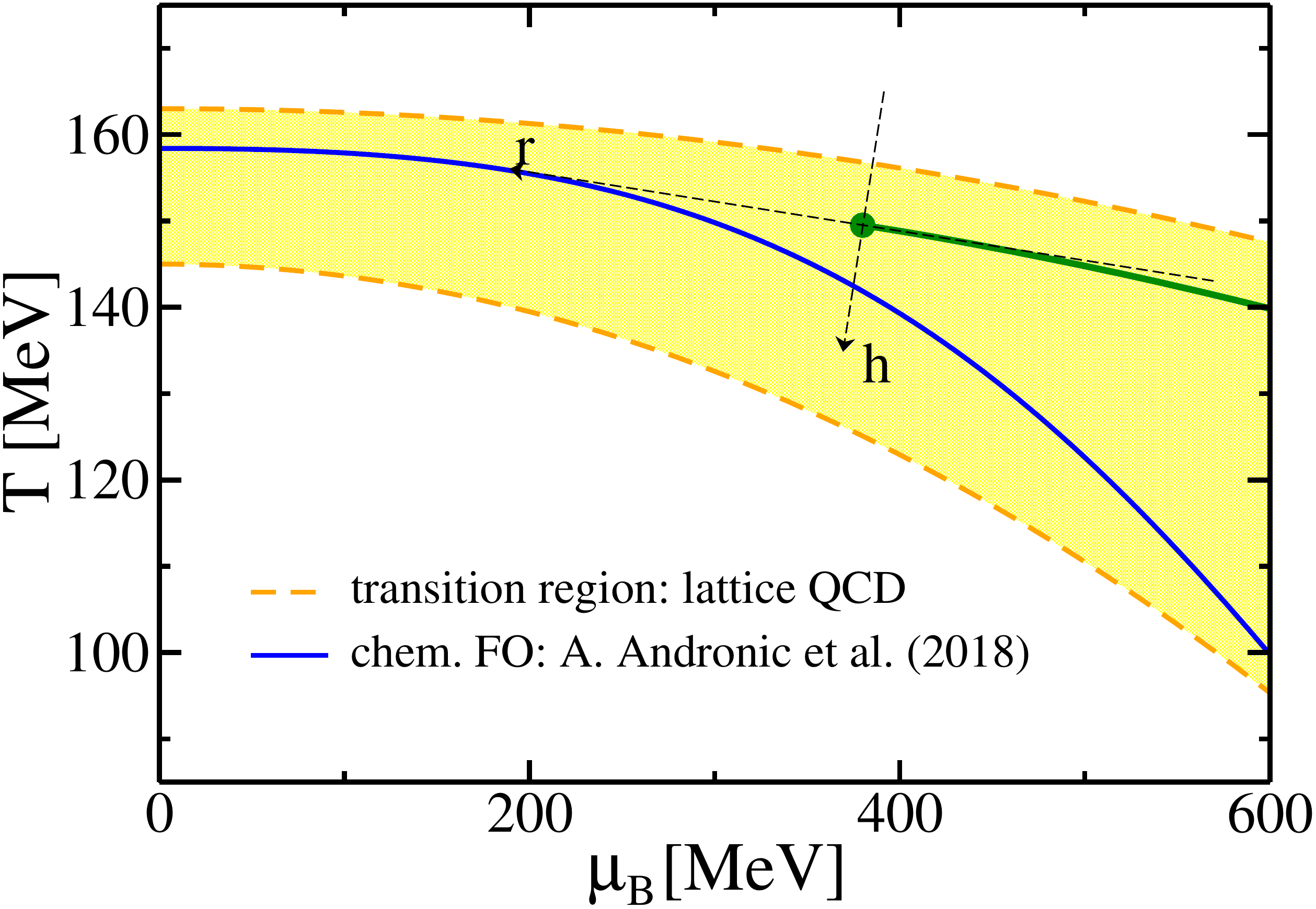}
\vskip-3mm\caption{(Color online) The model setup used in this work. The filled band between the two dashed curves shows lattice QCD constraints for the chiral crossover transition. The green dot denotes the critical point with the spin model coordinate system attached to it and the first-order phase transition line for larger baryon chemical potential. The solid blue line corresponds to the chemical freeze-out curve from~\cite{Andronic:2017pug}.}
\label{fig:setup}
\end{figure}

As a baseline model to calculate the net-proton number cumulants we choose the hadron resonance gas (HRG) model in which the number density of each particle species is given by the  ideal gas formula,
\begin{equation}
\label{eq:particledensity}
n_i(T,\mu_i)=d_i\int\frac{d^3k}{(2\pi)^3}f_i^0(T,\mu_i)\,.
\end{equation}
Here $d_i$ is the degeneracy factor and
\begin{equation}
\label{eq:distfunction}
f_i^0=\frac{1}{(-1)^{B_i}+e^{(E_i-\mu_i)/T}}
\end{equation}
is the equilibrium distribution function, where $E_i=\sqrt{\vec{p}^2+m_i^2}$ is the dispersion relation and $\mu_i=B_i\mu_B+S_i\mu_S+Q_i\mu_Q$ the chemical potential of a particle of mass $m_i$, baryon number $B_i$, strangeness $S_i$ and electric charge $Q_i$. $\mu_B$, $\mu_Q$ and $\mu_S$ denote baryon, strangeness and charge chemical potentials.

Since in HRG model the QCD pressure is approximated by a sum of partial ideal gas pressures corresponding to different particles, there are only thermal fluctuations in this approximation. To include critical fluctuations on top of the thermal ones we follow the phenomenological approach employed in Ref.~\cite{Bluhm:2016byc}. In this approach the particle mass is assumed to be composed of critical and non-critical parts as suggested in linear sigma models,
\begin{equation}
m_i\sim m_0+g_i\sigma\,,   
\end{equation}
where $m_0$ is a non-critical contribution and $g_i$ is the coupling strength between the critical mode and the particle of type $i$. Critical mode fluctuations modify the distribution function into $f_i= f_i^0+\delta f_i$, where the change of the distribution function due to critical mode fluctuations reads
\begin{equation}
\delta f_i=\frac{\partial f_i}{\partial m_i}\delta m_i=-\frac{g_i}{T}\frac{v_i^2}{\gamma_i}\delta\sigma \,,
\end{equation}
 with $v_i^2=f_i^0((-1)^{B_i}f_i^0+1)$ and $\gamma_i=E_i/m_i$.

Fluctuations of the particle number in the thermal medium can be quantified in terms of cumulants. The n-th order cumulant of i-th particle species reads
\begin{equation}
C^i_n=VT^3\frac{\partial^{n-1} (n_i/T^3)}{\partial(\mu_i/T)^{n-1} }\bigg\vert_T \,,
\end{equation}
where temperature $T$ is kept constant. In this work we consider the first four cumulants of the net-proton number, $N_{p-\bar{p}}=N_p-N_{\bar{p}}\,$, which are given by~\cite{Bluhm:2016byc}
\begin{eqnarray}\label{eq:c_n_original}
C_n&=&C_n^p+(-1)^nC_n^{\bar{p}}\\
&+&(-1)^n\langle(V\delta\sigma)^n\rangle_c (m_p)^n (J_p-J_{\bar{p}})^n\,,\nonumber
\end{eqnarray}
where $C_n^p$ and $C_n^{\bar{p}}$ are n-th order proton and anti-proton cumulants obtained within the baseline model, respectively, $\langle(V\delta\sigma)^n\rangle_c$ is n-th critical mode cumulant and
\begin{equation}
J_i=\frac{gd}{T}\int\frac{d^3k}{(2\pi)^3}\frac{1}{E}f_i^0(1-f_i^0)\,.
\label{eq:j_integral}
\end{equation}
Moreover, the contributions of other particles and resonance decays are neglected.

In general, cumulants of the critical mode cannot be calculated analytically. Following the approach introduced in Ref.~\cite{Bluhm:2016byc}, we model them using universality class arguments which state that different physical systems belonging to the same universality class exhibit the same critical behavior close to the critical point~\cite{Zinn-Justin}. Under the assumption that QCD belongs to the same universality class as the three-dimensional Ising model~\cite{Cp3,Cp4,Cp5} we can identify the QCD order parameter, $\sigma$, with the magnetization, $M_I$, the order parameter of the spin model. Hence, the critical mode cumulants can be written as~\cite{Bluhm:2016byc}
\begin{equation}
\langle(V\delta\sigma)^n\rangle_c=\left(\frac{T}{VH_0}\right)^{n-1}\left.\frac{\partial^{n-1}M_I}{\partial h^{n-1}}\right\vert_r\,,
\label{eq:m_cumulant}
\end{equation}
where $r=(T-T_c)/T_c$ is the reduced temperature and $h=H/H_0$ the reduced magnetic field. The critical point is located at $r=h=0$.

\begin{figure}
\vskip1mm
\includegraphics[width=0.85\linewidth]{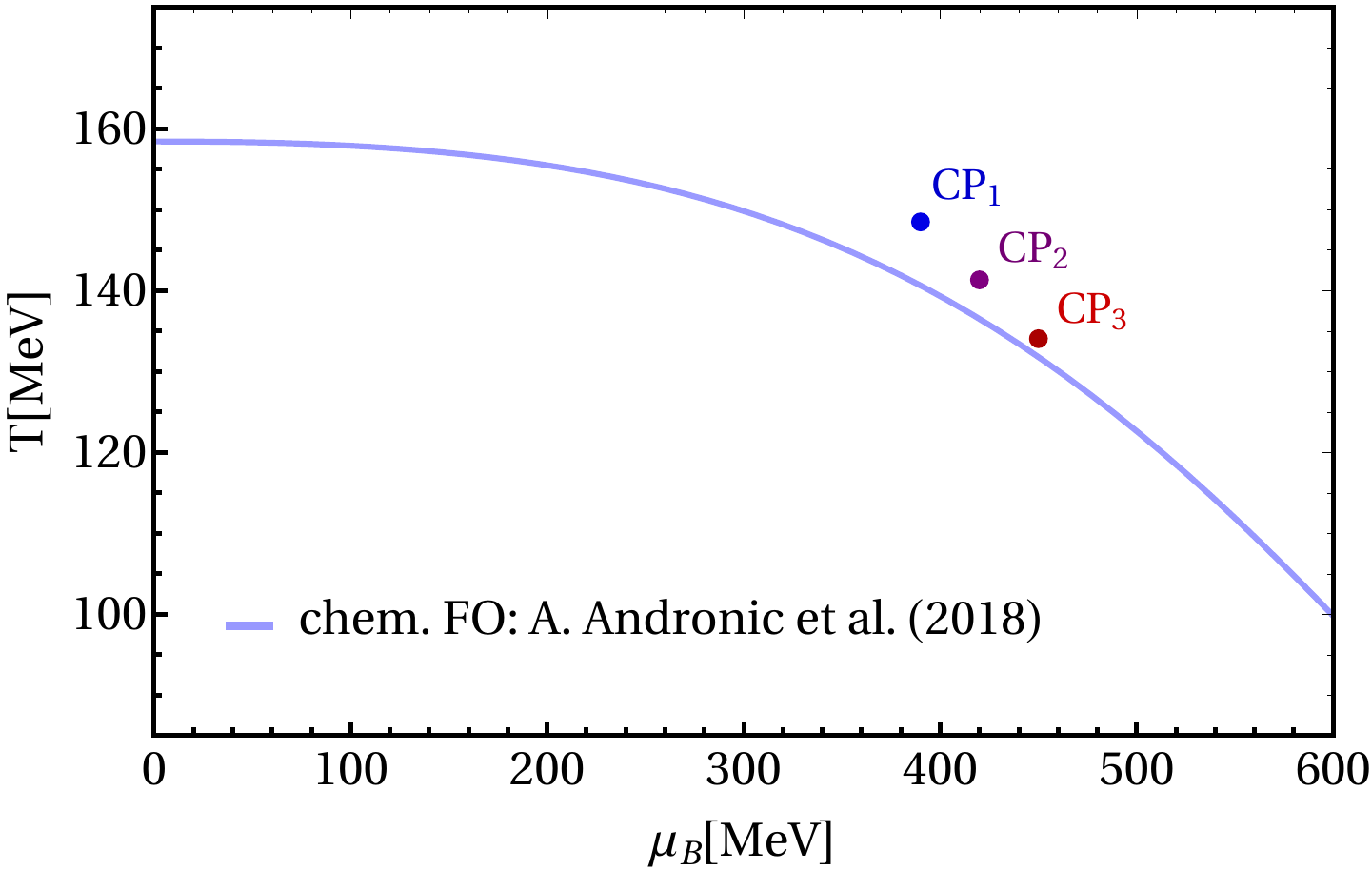}
\vskip-3mm\caption{(Color online) QCD critical point locations from Tab. \ref{tab:cp_setup} plotted with the chemical freeze-out curve~\cite{Andronic:2017pug} used in this work.}
\label{fig:cp_setup}
\end{figure}

\begin{table}[b]
\noindent\caption{Locations  of the QCD critical point in the ($\mu_B,T$)-plane considered in this work. These locations in the QCD phase diagram are shown in Fig.~\ref{fig:cp_setup}.}\vskip3mm\tabcolsep4.5pt
\noindent{\footnotesize
\begin{tabular}{|c|c|c|c|c|}
 \hline%
 \multicolumn{1}{|c}{\rule{0pt}{5mm} $CP_i$}%
 & \multicolumn{1}{|c|}{$\mu_{cp}\,$[MeV]}
 & \multicolumn{1}{|c|}{$T_{cp}\,$[MeV]}\\[2mm]%
\hline%
\rule{0pt}{5mm} 1 & 390 & 149 \\
\rule{0pt}{5mm} 2 & 420 & 141 \\
\rule{0pt}{5mm} 3 & 450 & 134 \\[2mm]%
\hline
\end{tabular}
}
\label{tab:cp_setup}
\end{table}

In the net-proton number cumulants, the singular part of the second cumulant receives a contribution from the first derivative of the order parameter with respect to the reduced magnetic field,
\begin{equation}
C_2^{\text{sing.}} \propto \frac{\partial M_I}{\partial h}.
\end{equation}
The right-hand side of this equation is the magnetic susceptibility of the Ising model which, due to universality, can be identified with the chiral susceptibility of QCD. The $C_2$, however, is related to the baryon number susceptibility which is known to diverge weaker than the chiral one~\cite{Hatta:2002sj,Sasaki:2006ws,Sasaki:2007qh}. Therefore, the model introduced in Ref.~\cite{Bluhm:2016byc} requires some modifications~\cite{Szymanski:2019yho}. This can be done using the following relation obtained within the effective model calculations on the mean field level~\cite{Sasaki:2006ws,Sasaki:2007qh},
\begin{equation}
\chi_{\mu\mu}\simeq\chi_{\mu\mu}^{reg}+\mathbf{\sigma}^2\chi_{\text{chiral}}\,,
\end{equation}
in which the singular contribution to the baryon number susceptibility is proportional to the chiral susceptibility times the squared order parameter and $\chi_{\mu\mu}^{reg}$ is the regular part of the baryon number susceptibility. To obtain such a form of the second cumulant, the proton mass in Eq. \eqref{eq:c_n_original} should be replaced by the order parameter, $\sigma$, such that the new $C_2$ reads
\begin{equation}
C_2=C_2^p+C_2^{\bar{p}}+g^2\sigma^2 \langle(V\delta\sigma)^n\rangle (J_p-J_{\bar{p}})^2\,.
\label{eq:c2_model}
\end{equation}
The modified higher order cumulants are
\begin{equation}
C_3=C_3^p-C_3^{\bar{p}}- g^3\sigma^3 \langle(V\delta\sigma)^n\rangle (J_p-J_{\bar{p}})^3    
\label{eq:c3_model}
\end{equation}
and
\begin{equation}
C_4=C_4^p+C_4^{\bar{p}}+g^4\sigma^4 \langle(V\delta\sigma)^n\rangle (J_p-J_{\bar{p}})^4\,.
\label{eq:c4_model}
\end{equation}
Since the cumulants are volume-dependent it is convenient to consider their ratios in which this dependence cancels out,
\begin{equation}
\frac{C_2}{C_1}=\frac{\sigma^2}{M}\,,\qquad \frac{C_3}{C_2}=S\sigma\,,\qquad \frac{C_4}{C_2}=\kappa\sigma^2\,,
\label{eq:ratios}
\end{equation}
where $M=C_1$ is the mean, $\sigma^2=C_2$ the variance, $\kappa=C_4/C_2^2$ the kurtosis and $S=C_3/C_2^{3/2}$ the skewness.

To use universality class arguments discussed above, a mapping between the QCD phase diagram and reduced temperature and magnetic field of the spin model is needed. Such a mapping is non-universal and has to be modeled for each system separately. In this work we use a linear mapping~\cite{Mukherjee:2015swa,Nonaka:2004pg} in which the critical point is located at $r=h=0$, the $r$ axis is tangential to the QCD first order phase transition line and the positive direction of the $h$ axis points towards the hadronic phase. Schematically, this is shown in Fig. \ref{fig:setup}, where the green line denotes the first order phase transition and the filled band shows lattice QCD constraints on the location of the chiral crossover region.

To calculate the order parameter as well as its cumulants we use the parametric representation of the magnetic equation of state~\cite{Guida:1996ep}. For a more detailed discussion of the mapping, lattice limits as well as the magnetic equation of state we refer the reader to the papers~\cite{Bluhm:2016byc,Szymanski:2019yho}.

Finally, assuming that the matter created during a heavy ion collision forms a thermal medium characterized by temperature and chemical potentials, experimental data on event-by-event multiplicity fluctuations can be compared with model results. To this end, we calculate the net-proton number cumulants at the chemical freeze-out. The chemical freeze-out conditions used in this work were obtained by the analysis of hadron yields~\cite{Abelev:2013vea,Abelev:2013xaa,ABELEV:2013zaa,Abelev:2014uua,Adam:2015yta,Adam:2015vda}. The blue line in Fig. \ref{fig:setup} shows the recently obtained parametrization~\cite{Andronic:2017pug}.

\begin{figure}
\vskip1mm
\includegraphics[width=0.85\linewidth]{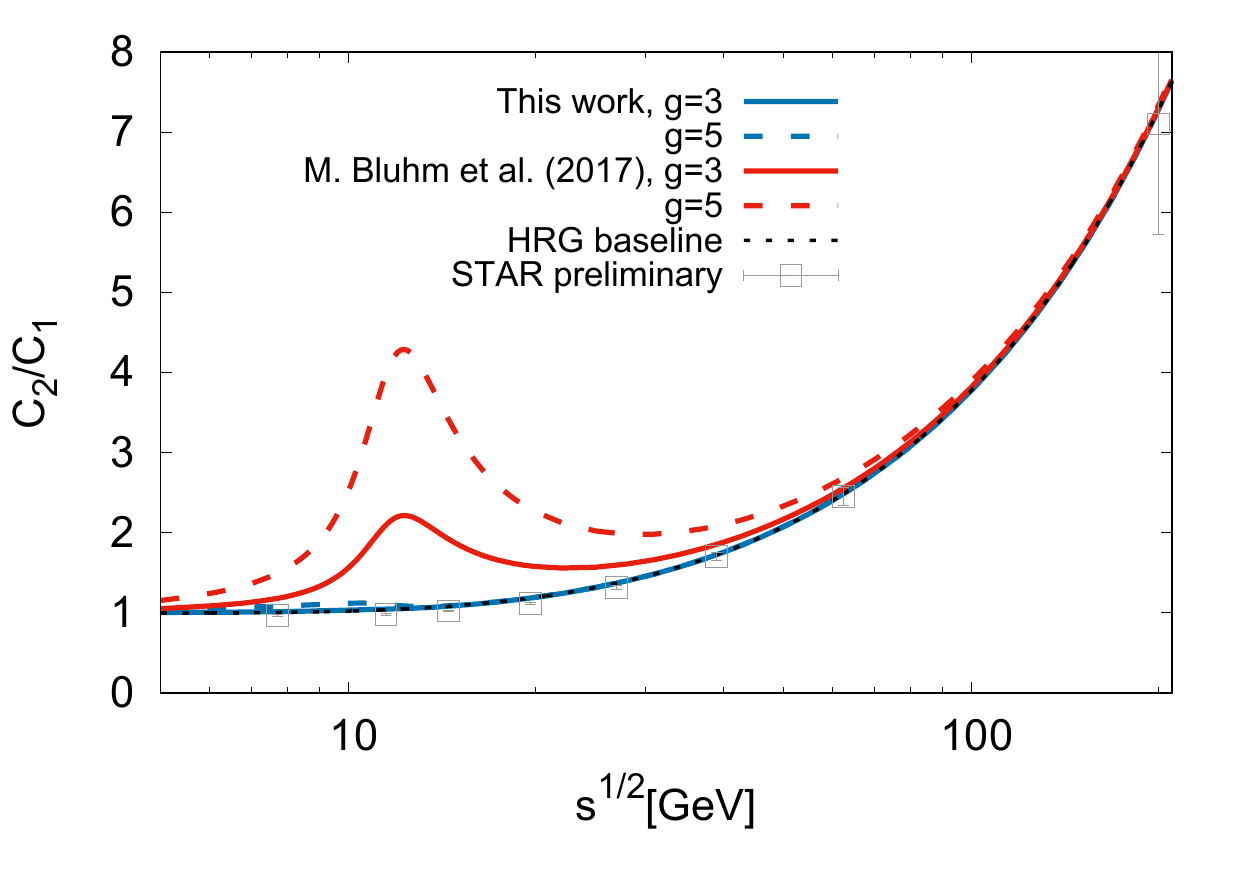}
\vskip-3mm\caption{(Color online) The second to first net-proton number cumulant ratio for $g=3$ and $5$ calculated following Ref.~\cite{Bluhm:2016byc} (red solid and dashed lines, respectively) compared to refined model results~\cite{Szymanski:2019yho} (blue solid and dashed lines, respectively). The preliminary STAR data on the net-proton number fluctuations~\cite{Thader:2016gpa} (squares with the error bars containing both statistical and systematic errors) and HRG baseline result (black dotted line) are also shown for comparison.}
\label{fig:imp_vs_orig}
\end{figure}

\begin{figure}
\vskip1mm
\includegraphics[width=0.85\linewidth]{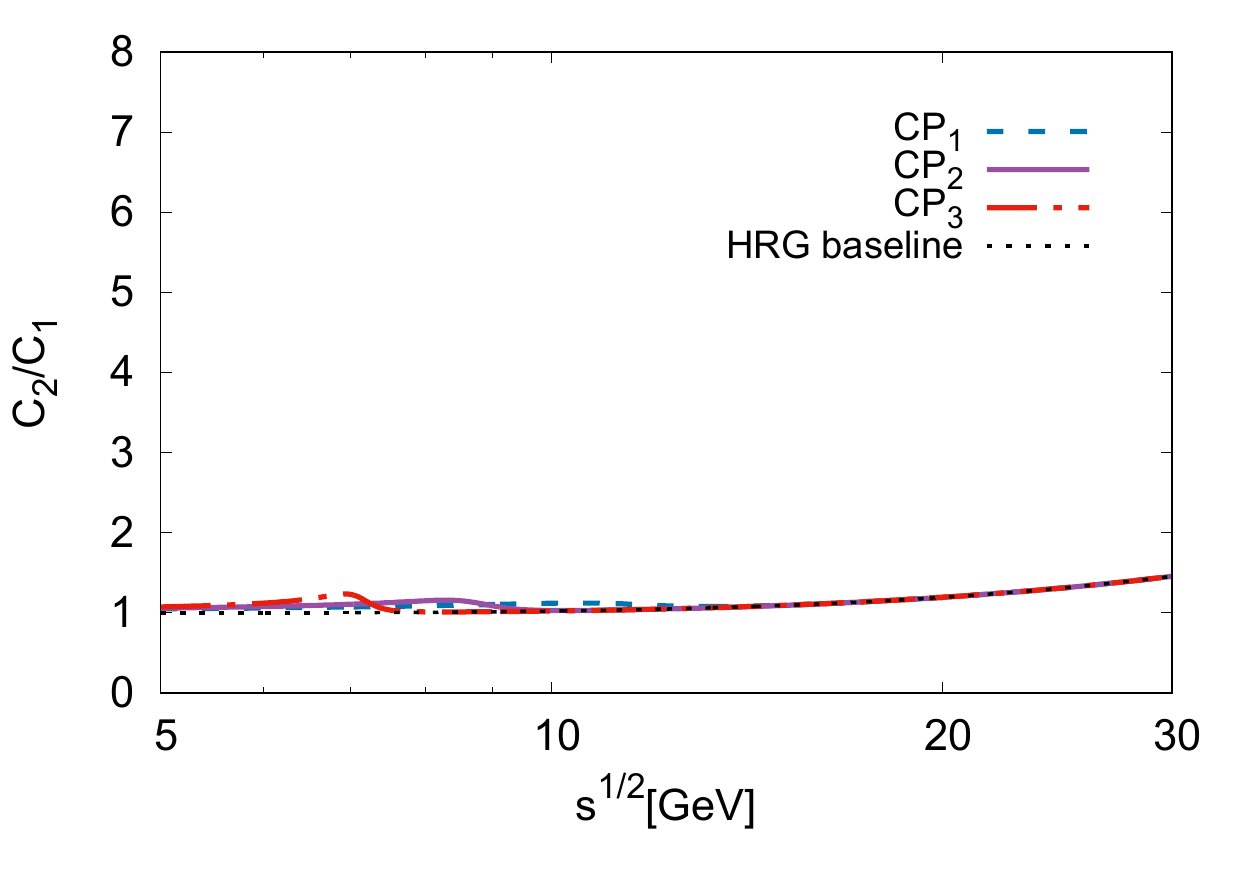}\\
\includegraphics[width=0.85\linewidth]{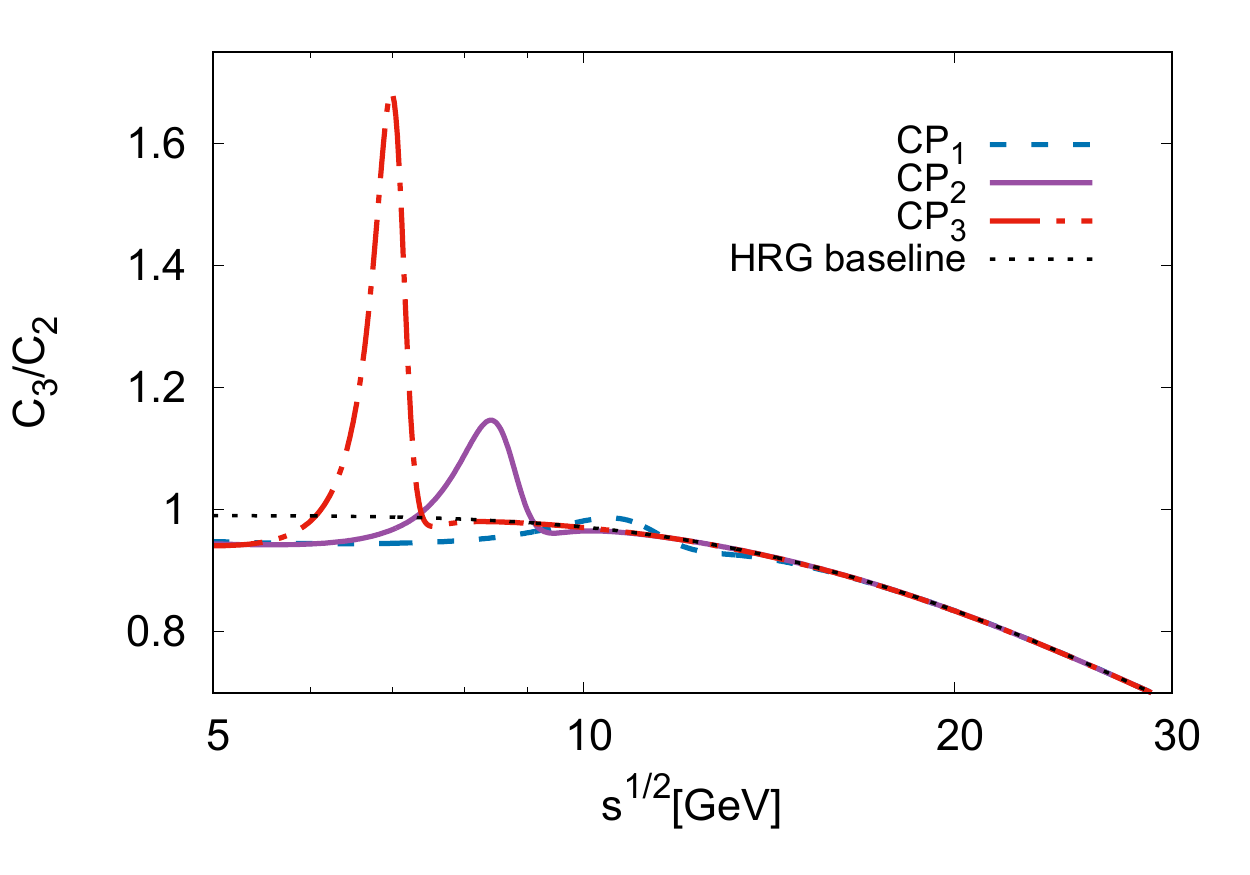}\\
\includegraphics[width=0.85\linewidth]{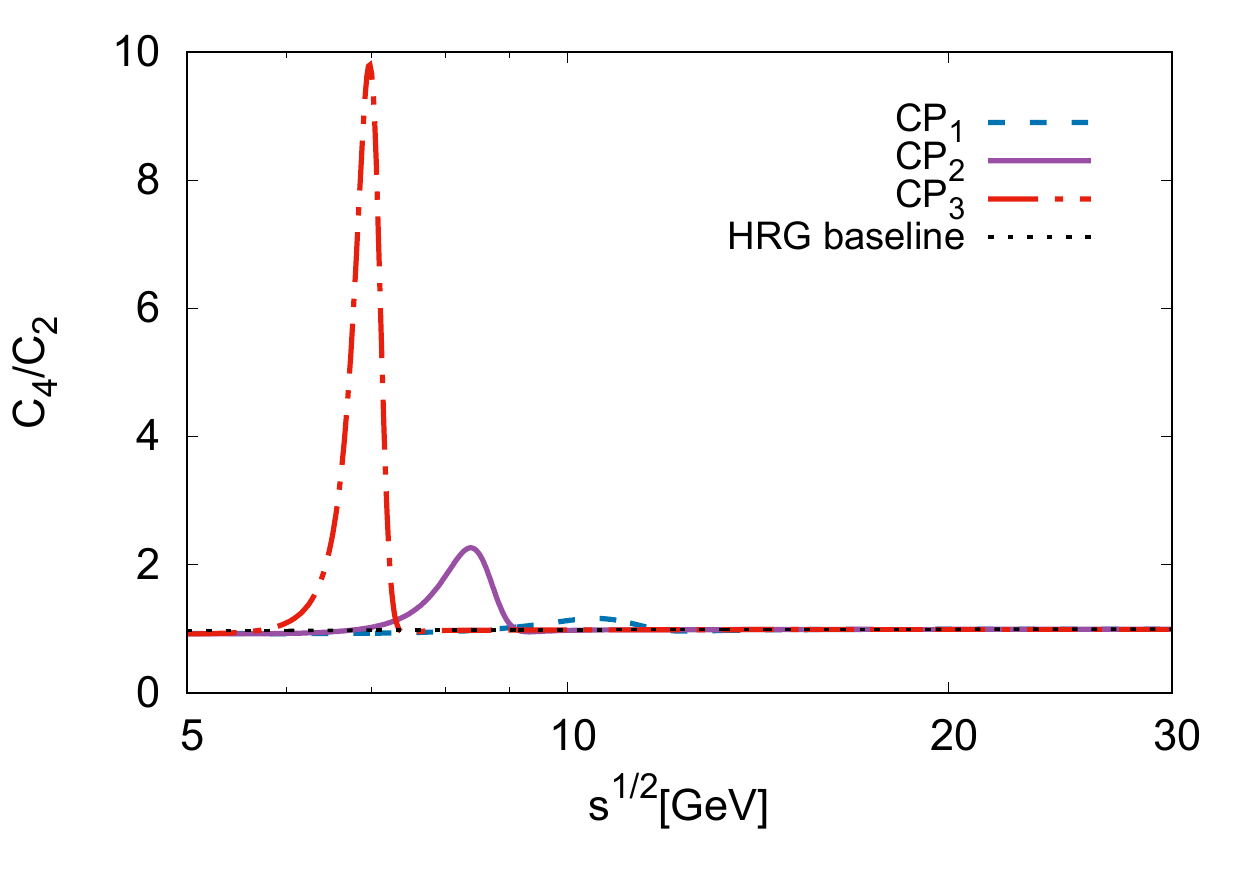}
\vskip-3mm\caption{(Color online) Ratios of net-proton number cumulants calculated in the refined model~\cite{Szymanski:2019yho} for fixed coupling $g=5$ and for different locations of the QCD critical point (listed in Tab. \ref{tab:cp_setup}).
}
\label{fig:cp_g_5}
\end{figure}

\begin{figure}
\vskip1mm
\includegraphics[width=0.85\linewidth]{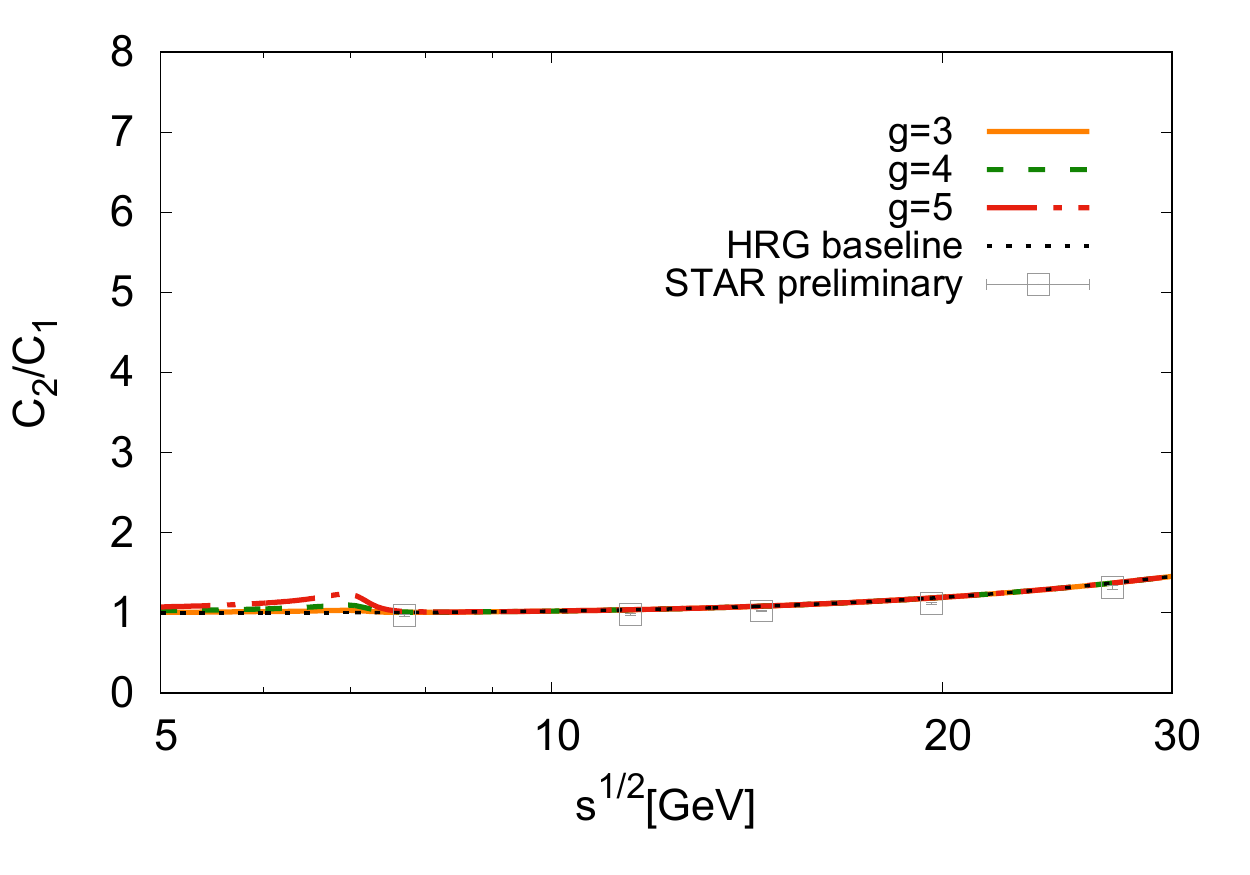}\\
\includegraphics[width=0.85\linewidth]{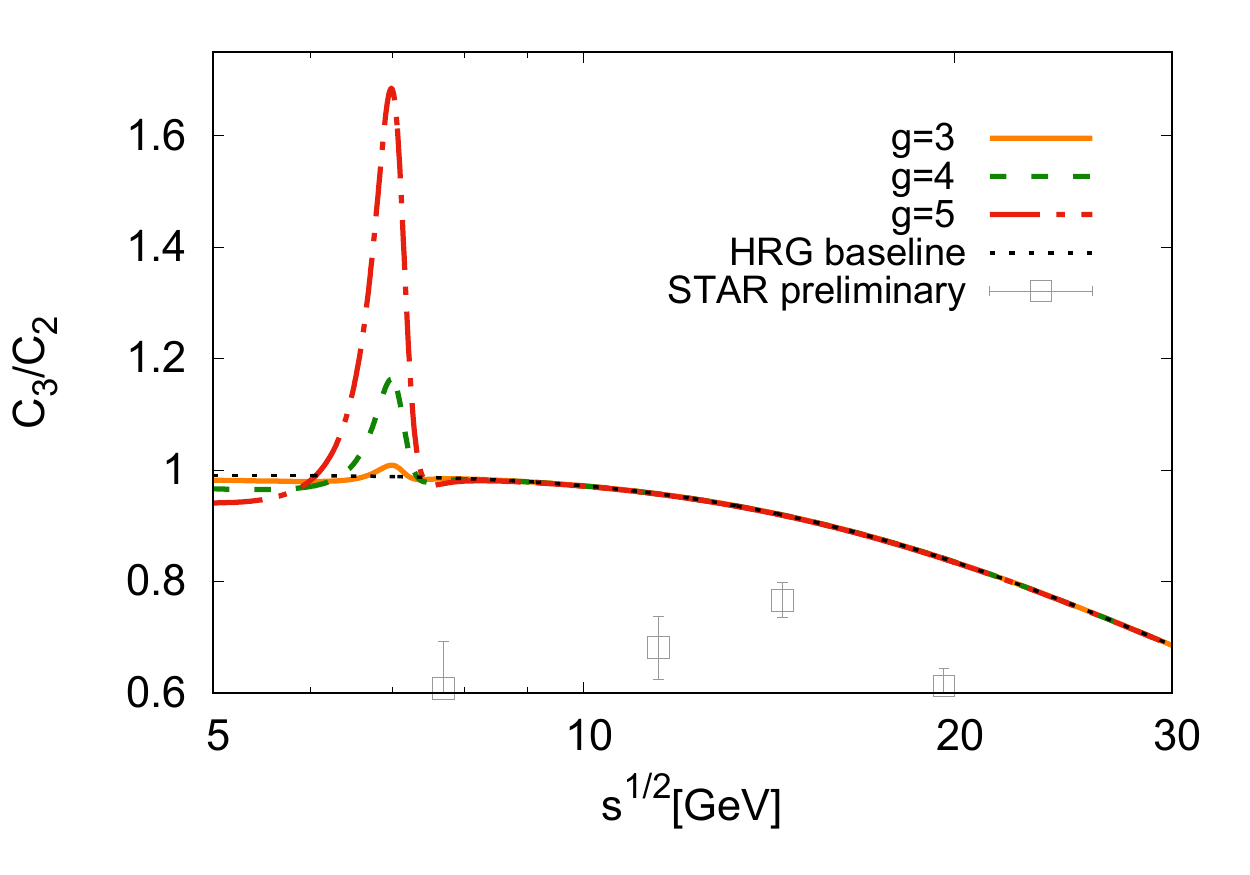}\\
\includegraphics[width=0.85\linewidth]{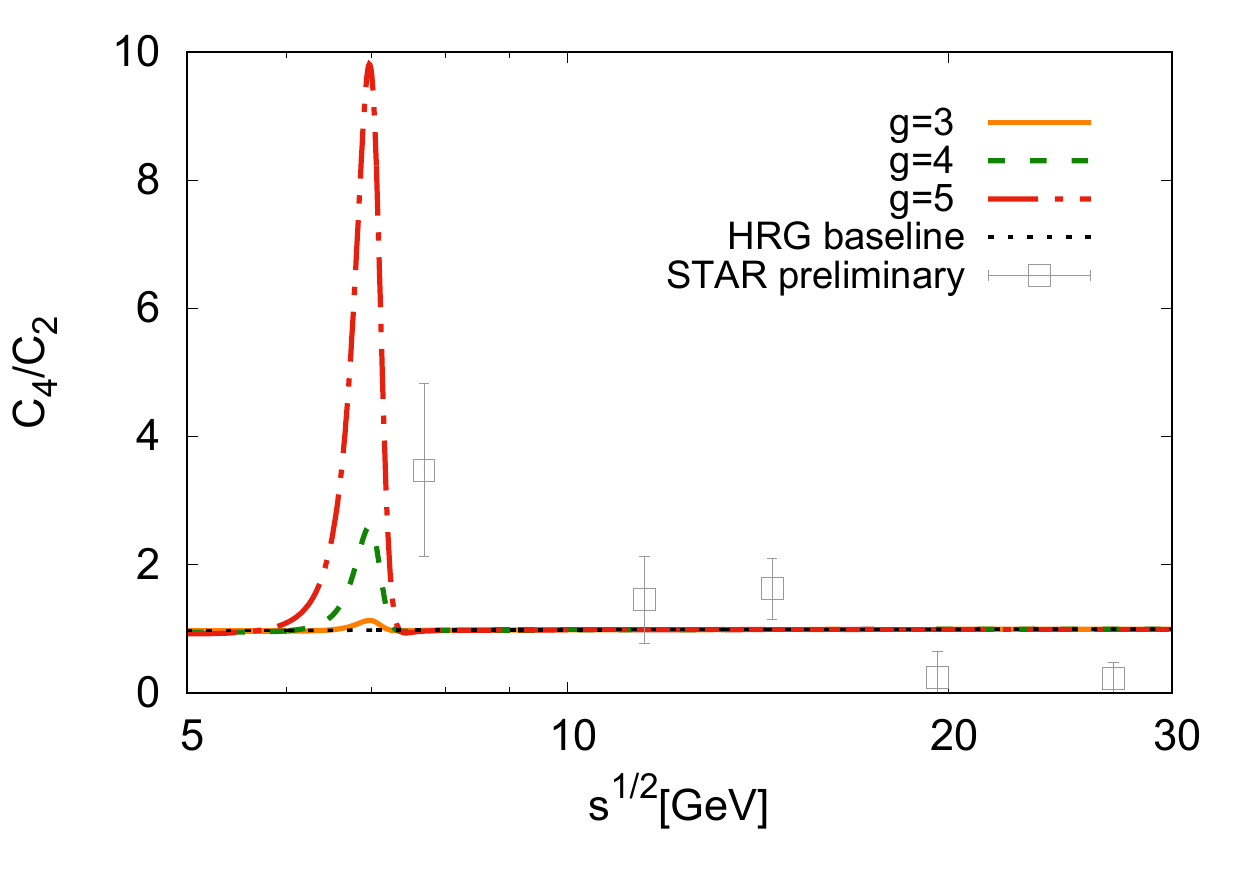}
\vskip-3mm\caption{(Color online)  Ratios of net-proton number cumulants calculated in the refined model~\cite{Szymanski:2019yho} with CP$_3$ and for coupling strengths, $g=$ 3, 4 and 5 (orange solid, green long-dashed and red dash-dotted lines, respectively). The preliminary STAR data on the net-proton number fluctuations~\cite{Thader:2016gpa} (squares with the error bars containing both statistical and systematic errors) and HRG baseline results (black dotted lines) are also shown for comparison.}
\label{fig:cp_3_g_comparison}
\end{figure}

\section{Numerical results}

In this section we discuss numerical results on net-proton number cumulant ratios obtained within the current model. The set of model parameters includes the coupling strength $g$ between (anti)protons and the critical mode, the parameters of the magnetic equation of state as well as the size of the critical region in the $(T,\mu)$ plane. Their values as well as a detailed discussion can be found in Refs.~\cite{Bluhm:2016byc,Szymanski:2019yho}. Moreover, the location of the QCD critical point is unknown. To study the effect of its position in the QCD phase diagram on the refined model results, we consider three different locations of the CP listed in Tab. \ref{tab:cp_setup} and shown in Fig. \ref{fig:cp_setup}, where the distance to the freeze-out curve is the farthest for CP$_1$ and closest for $CP_3$. 

The first step of our discussion is the comparison between the $C_2/C_1$ ratio obtained using the model from Ref.~\cite{Bluhm:2016byc}, where the n-th net-proton number cumulant is given by Eq. \eqref{eq:c_n_original}, and the refined model~\cite{Szymanski:2019yho} for the critical point location CP$_1$. This is shown in 
Fig.~\ref{fig:imp_vs_orig}. Results obtained using the original model exhibit clear non-monotonic behavior and deviate strongly from the non-critical baseline (the black-dotted line)  even for small coupling, $g= 3$, which becomes more pronounced for $g=5$ (the red solid and dashed lines, respectively). Using the current approach we find a substantial reduction of criticality in $C_2/C_1$ ratio, even for larger values of $g$ (see the blue curves in Fig. \ref{fig:imp_vs_orig}). The refined model results for the $C_2/C_1$ ratio agree with the experimental data from the STAR Collaboration~\cite{Thader:2016gpa}. On the other hand, the original model would require exceptionally small coupling strength in order to capture the experimentally observed behavior.

The net-proton cumulant ratios obtained in the refined model for different locations of the critical point (as listed in Tab. \ref{tab:cp_setup}) and with a fixed value of coupling, $g=5$, are shown in Fig. \ref{fig:cp_g_5}. We find that a non-monotonic behavior of cumulant ratios becomes more pronounced when the critical point is closer to the freeze-out line. Moreover, the deviation from the non-critical HRG baseline becomes larger for higher order cumulant ratios.

Finally, Fig.~\ref{fig:cp_3_g_comparison} shows the coupling strength dependence of net-proton number cumulant ratios obtained for CP$_3$. We find a strong $g$ dependence of all ratios which is expected since in our refined model the n-th cumulant scales as $g^{2n}$, according to Eqs. \eqref{eq:j_integral} and \eqref{eq:c2_model}-\eqref{eq:c4_model}. When our model results are compared to the STAR data~\cite{Thader:2016gpa}, we find a qualitative agreement with the $C_2/C_1$ and $C_4/C_2$ ratios. On the other hand, the $C_3/C_2$ ratio does not follow the systematics seen in the data, i.e. our model results overshoot the HRG baseline while the data stay below.

Our results suggest that the appropriate choice of model parameters as well as the location of the QCD critical point allows us to describe some of the experimentally observed cumulant ratios. Especially, the smooth dependence of $C_2/C_1$ and strong increase of $C_4/C_2$ at low beam energies, $\sqrt{s}<20\,$GeV, seen by the STAR Collaboration, suggest that the QCD critical point may be located close to the freeze-out curve. However, in this case the $C_3/C_2$ ratio should increase beyond the non-critical baseline, which is not seen in the experimental data. Therefore it seems unlikely that the QCD critical point is close to the freeze-out curve. This conclusion, however, requires additional theoretical and experimental justifications due to uncertainties in the model parameters as well as in the experimental data.

\section{Conclusions}

We presented ratios of net-proton number cumulants obtained within an effective model in which the coupling between (anti)protons and critical mode fluctuations is introduced by connecting particle masses to the order parameter. We modified the existing approach~\cite{Bluhm:2016byc} to take into account the correct scaling properties of the baryon number susceptibility as dictated by the universality hypothesis.

Model results were compared with the recent experimental data on net-proton number fluctuations from the STAR Collaboration. We find a substantial reduction of the signal coming from the presence of the QCD critical point in the $C_2/C_1$ ratio which stays in agreement with the experimental data. Moreover, we find that the model discussed in the present work allows us to describe some of the experimentally observed features in the net-proton number cumulant ratios. Particularly, smooth dependence of $C_2/C_1$ and increase of $C_4/C_2$ at lower beam energies ($\sqrt{s}<20\,$GeV) suggest that the critical point may be located close to the freeze-out curve. However, the experimentally observed $C_3/C_2$ ratio does not follow the behavior expected from such a scenario. 

Therefore, it seems unlikely that the QCD critical point is located close to the phenomenological freeze-out curve. However, because of uncertainties on both theoretical as well as experimental sides, this statement requires further investigation.

\vskip3mm \textit{M.~S. and M.~B. acknowledge support of the  Polish National Science Center (NCN) under grant Polonez UMO-2016/21/P/ST2/04035. M.~B. was partly supported by the program "Etoiles montantes en Pays de la Loire 2017". K.~R. and C.~S. were supported in part by the Polish National Science Center (NCN) under Maestro grant DEC-2013/10/A/ST2/00106. K.~R. also acknowledges partial support of the Polish Ministry of Science and Higher Education. We thank F.~Geurts and J.~Th\"ader for providing the preliminary STAR data~\cite{Thader:2016gpa} shown in Figures~\ref{fig:imp_vs_orig} and~\ref{fig:cp_3_g_comparison}.
}

\begin{flushright}
{\footnotesize Received 18.07.19}
\end{flushright}

\end{document}